\documentclass[twocolumn,showpacs,floatfix,aps,pra]{revtex4}

\usepackage[dvips]{graphicx,color}% Include figure files
\usepackage{dcolumn}% Align table columns on decimal point
\usepackage{bm}% bold math
\usepackage{amsmath, amssymb, mathrsfs}

\newcommand{\rr}{\mathbf{r}}
\newcommand{\kk}{\mathbf{k}}
\newcommand{\KK}{\mathbf{K}}
\newcommand{\QQ}{\mathbf{Q}}
\newcommand{\qq}{\mathbf{q}}
\newcommand{\RR}{\mathbf{R}}
\newcommand{\be}{\begin{equation}}
\newcommand{\ee}{\end{equation}}
\newcommand{\bea}{\begin{eqnarray}}
\newcommand{\eea}{\end{eqnarray}}
\def\epsb{\mbox{\boldmath$\epsilon$}}
\def\eps{\mbox{$\epsilon$}}
\def\epsbs{\mbox{\boldmath\scriptsize$\epsilon$}}
\def\cheta{\tilde}

\begin{document}
\title{Fano-Hopfield model and photonic band gaps for an arbitrary 
atomic lattice}

\author{Mauro Antezza}
\affiliation{Laboratoire Kastler Brossel, \'{E}cole Normale Sup\'{e}rieure, 
CNRS and UPMC, 24 rue Lhomond, 75231 Paris, France}
\author{Yvan Castin}
\affiliation{Laboratoire Kastler Brossel, \'{E}cole Normale Sup\'{e}rieure, 
CNRS and UPMC, 24 rue Lhomond, 75231 Paris, France}

\date{\today}

\begin{abstract}
We study the light dispersion relation in a periodic ensemble of atoms at fixed positions
in the Fano-Hopfield model (the atomic dipole being modeled with harmonic oscillators).
Compared to earlier works, we do not restrict to cubic
lattices, and we do not regularize the theory by hand but we renormalize it in a systematic
way using a Gaussian cut-off in momentum space.
Whereas no omnidirectional spectral gap is known for light in a
Bravais atomic lattice, 
we find that, for a wide range of parameters, an omnidirectional gap occurs in a diamond atomic lattice,
which may be realized in an experiment with ultra-cold atoms.
The long-wavelength limit of the theory also provides a Lorentz-Lorenz (or Clausius-Mossotti)
relation for an arbitrary lattice.
\end{abstract}

\pacs{42.50.Ct, 67.85.-d, 71.36.+c}
% light matter interaction ; ultracold gases, trapped gases ; polaritons

\maketitle

\section{Introduction}

The determination of the spectrum of light in a periodic ensemble of atoms is a fundamental
problem that still raises intriguing questions.
After the seminal work of Hopfield \cite{Hopfield58}
(see also Agranovich \cite{Agranovich60}), based on the Fano oscillator
model for the atomic dipole \cite{Fano56},
many theoretical works have been performed \cite{LagendijkRMP}.
Most of them, inspired by the typical context of condensed matter physics 
considered in \cite{Hopfield58},
focus on the long-wavelength limit where Lorentz-Lorenz (or Clausius-Mossotti)
type relations may be derived \cite{Juzeliunas07,Carusotto08}.
Recently, this problem was extended to the whole Brillouin zone in the case of cubic lattices
\cite{Lagendijk96,Knoester06}, which allows to address the presence or the absence
of an omnidirectional spectral gap for light.

This problem of light spectrum in atomic lattices is no longer a purely theoretical issue.
Recent experiments with ultracold atoms, having led to the observation of a Mott
phase with one atom per lattice site \cite{Bloch02}, have indeed opened up the possibility
to investigate the propagation of light in an atomic lattice,
taking advantage of the large variety of optical lattices that may be realized
to trap the atoms \cite{Grynberg94}, all this in the regime where the lattice spacing 
is of the order of the optical wavelength, so that a probing of the whole
 Brillouin zone can be envisaged.

In the photonic crystals in solid state systems, made of dielectric spatially
extended objects (rather than point-like atoms), after the pioneering work
of \cite{Soukoulis90} for diamond lattices of dielectric spheres,
many configurations are now known to lead to a spectral gap, with a variety of applications to light
trapping and guiding \cite{Yablo91,FCBook}.
On the contrary, for atomic lattices, 
no omnidirectional spectral gap was found, neither in cubic atomic lattices \cite{Knoester06}
nor in several less symmetric Bravais lattices \cite{Castin09}.

Two factors determine the presence of an omnidirectional gap: crystal geometry and details of the 
light-matter scattering process, both can separately close a gap. Indeed, in photonic crystals 
materials, characterized by a macroscopic modulation of the refractive index, the same lattice geometry can 
lead or not to an omnidirectional gap depending on the
modulation of the refractive index, as is the case for the simple cubic (sc) or body centered cubic (bcc) 
 lattices \cite{John04}. Differently from photonic crystals materials, 
in the physics of ultracold atoms, 
one can realize periodic structures with a single atom per site 
\cite{Bloch02} rather than a macroscopic number,
realizing an ideal crystalline structure with a variety of possible geometries
\cite{Grynberg94}. Atoms are scatterers characterized by 
 a strong resonant and point-like interaction with light, so that the features of the
 light propagation in an atomic lattice 
cannot be straightforwardly extrapolated from known results in
solid state photonic crystals.
Also in the atomic case, it is not  possible, only by geometric 
 considerations, to predict the presence of an omnidirectional photonic band gap,
the details of the light-matter scattering process do matter
\cite{Knoester06}.

Here we develop, within the Fano-Hopfield model,
a self-consistent 
theory for the elementary excitation spectrum of the light-atom 
field in atomic periodic systems, 
which is valid not only for 
cubic symmetry atomic lattices \cite{Knoester06}  
but also for any Bravais lattice, and, 
even more, also for periodic non-Bravais lattices (i.e.\ for crystals
with several atoms per primitive unit cell).
Our theory includes the light polarizations degrees of freedom, and is based on the 
introduction of a Gaussian momentum cut-off which allows to eliminate
all divergences (even in a periodic infinite system of atoms) by a systematic
renormalization procedure.
We then use our theory to address the existence of an omnidirectional spectral
gap for light in atomic lattices.
In particular, we show that the diamond atomic lattice, 
which is a non-Bravais lattice composed of two identical face-centered
cubic (fcc) atomic lattices deduced
one from the other by a translation along the main diagonal of the cube,
may support an omnidirectional gap for light.
To our knowledge, this is the first example of the occurrence of such a gap 
in a periodic structure of point-like atomic scatterers.

The paper is organized as follows. We present the model Hamiltonian in section \ref{sec:model}.
We renormalize the model and obtain an implicit equation for the light spectrum,
first for Bravais lattices in section \ref{sec:bravais} (with a comparison
to existing predictions for a fcc lattice),
and then for a general lattice in section \ref{sec:nonBravais}.
We derive Lorentz-Lorenz relations in the long wavelength limit for an arbitrary
lattice in section \ref{sec:lorentz}.
In section \ref{sec:diamant} we calculate the spectrum in a diamond atomic lattice,
we discuss the existence of a gap and experimental issues such as the practical
realization of a diamond atomic lattice. 
We conclude in section \ref{sec:conclusion}.

\section{The model}   
\label{sec:model}

We consider $N$ atoms with a dipolar coupling to the electromagnetic field,
each atom being modeled by harmonic oscillators in the spirit of the Fano-Hopfield
model \cite{Fano56,Hopfield58}.
The atoms have fixed positions on a Bravais lattice (in section \ref{sec:nonBravais}
we will consider the even more general case of non-Bravais lattices). 
The $i^{\rm th}$ atom is in lattice site $\RR_i$. 
The Hamiltonian may be written as
\begin{eqnarray}
H &=& \sum_{i=1}^{N} \sum_{\alpha}^{x,y,z} \hbar \omega_B \hat{b}_{i,\alpha}^\dagger
\hat{b}_{i,\alpha}
+\sum_{\kk\in\mathcal{D}} \sum_{\epsbs\perp\kk} \hbar c k\, \hat{a}^\dagger_{\kk\epsbs} \hat{a}_{\kk\epsbs}\notag \\
&-&\sum_{i=1}^{N} \hat{\mathbf{D}}_i \cdot  \hat{{\mathbb E}}_{\perp}(\RR_i).
\label{eq:H}
\end{eqnarray}
The first two terms in (\ref{eq:H}) correspond to the uncoupled atomic and radiation 
contributions respectively, while the last term is the dipolar coupling operator. 
The operators $\hat{b}_{i,\alpha}$ and $\hat{b}_{i,\alpha}^\dagger$
respectively destroy and create an atomic excitation for the atomic dipole $i$ 
along spatial direction $\alpha$.
They obey the \emph{bosonic} commutation relations $[\hat{b}_{i,\alpha},\hat{b}^\dagger_{j,\beta}]=
\delta_{ij} \delta_{\alpha\beta} $ and $[\hat{b}_{i,\alpha},\hat{b}_{j,\beta}]=0$. 
In (\ref{eq:H}), $\omega_B$ is the \emph{bare} atomic resonance frequency, 
the sum $\sum_\alpha$ over the three directions of space $x$, $y$, $z$ 
accounts for the three spatial components of the dipoles.

The photon annihilation and creation operators also obey usual bosonic commutation 
relations
such as $[\hat{a}_{\kk\epsbs},\hat{a}^\dagger_{\kk'\epsbs'}]=
\delta_{\epsbs\epsbs'} \delta_{\kk\kk'}$ and $[\hat{a}_{\kk\epsbs},\hat{a}_{\kk'\epsbs'}]=0$,
where $\epsb$ and $\kk$ are the photon polarization and wavevector.
$c$ is the velocity of light in vacuum.
We assume a quantization volume $\textrm{V}$ which includes $N=\prod_{\gamma} M_\gamma$ 
atoms and corresponds
to periodic boundary conditions of the field,
with a period $M_{\gamma} {\bf e}_{\gamma}$ along each  direction $\gamma\in \{1,2,3\}$ of
the Bravais lattice,
with $M_{\gamma}\in\mathbb{N}$ and  ${\bf e}_{\gamma}$ is one of the three basis vectors 
of the Bravais lattice in direct space.
As the consequence, the allowed wavevectors $\kk$ for the electromagnetic field
belong to the discrete set 
$\mathcal{D}=\{\kk \;|\; 
\kk=\sum_\gamma (m_{\gamma}/M_\gamma)\, \tilde{{\bf e}}_{\gamma}, 
\forall m_{\gamma}\in\mathbb{Z} \}$, 
where  the
$\tilde{{\bf e}}_{\gamma}$'s are basis vectors of the reciprocal lattice,
such that 
$\tilde{\mathbf{e}}_\gamma \cdot \mathbf{e}_{\gamma'} = 2\pi \delta_{\gamma\gamma'}$.
The atom-electromagnetic field coupling term involves the dipole operator $\hat{\mathbf{D}}_i$
of the $i^{\textrm{th}}$ atom, whose component along direction $\alpha$
\be
\hat{D}_{i,\alpha} =  
\textrm{d}_B \left[ \hat{b}_{i,\alpha} + \hat{b}_{i,\alpha}^\dagger \right]
\ee
is proportional to the \emph{bare} atomic dipole moment $\textrm{d}_B$, and 
to the 
operator $\hat{{\mathbb E}}_{\perp}(\rr)$ 
\be
\hat{{\mathbb E}}_{\perp}(\rr) =\int d^3u\;\hat{{\bf E}}_{\perp}(\rr-\mathbf{u})
\;\chi(\mathbf{u};b),
\label{conv}
\ee
which is the convolution of the 
transverse electric field operator
\be
\hat{{\bf E}}_{\perp}(\rr) =
\sum_{\kk\in\mathcal{D}} \sum_{\epsbs\perp\kk} \left[\mathcal{E}_k
\epsb\; \hat{a}_{\kk\epsbs}\; e^{i\kk\cdot \rr} + \textrm{h.c.}\right]
\label{efield}
\ee
with a normalized cut-off function $\chi(\rr;b)$, 
$\int d^3r\;\chi(\rr;b)=1$, where  
$\mathcal{E}_k=i [\hbar kc/(2\varepsilon_0\textrm{V})]^{1/2}$,
 and the length $b$ is the cut-off parameter. 
Here $\chi(\rr;b)$ is a real and even function of $\rr$.
This convolution regularizes the theory by eliminating ultraviolet
divergences; it makes the model 
Hamiltonian $H$ well defined.  From (\ref{conv}) and (\ref{efield}) one has that 
\be
\hat{{\mathbb E}}_{\perp}(\rr) 
=\sum_{\kk\in\mathcal{D}} \sum_{\epsbs\perp\kk} \left[\mathcal{E}_k
\epsb\; \hat{a}_{\kk\epsbs}\; e^{i\kk\cdot \rr} + \textrm{h.c.} \right]\cheta{\chi}(\kk;b),
\ee
where the Fourier transform of $\chi(\rr;b)$ with respect to $\rr$,
\be
\cheta{\chi}(\kk;b)=\int d^3r\; e^{-i\kk\cdot\rr}\;\chi(\rr;b),
\ee
is a real and even function of $\kk$.

\section{Light dispersion relation in a Bravais lattice} 
\label{sec:bravais}

Starting from the Hamiltonian (\ref{eq:H}) it is possible 
(as done in Appendix \ref{app:exspect}) 
to derive the equations of motion for the matter and light fields, and, after a linear 
transformation and the use of the Bloch theorem, furthermore leaving out
the so-called {\sl free-field} solutions (see below),
one obtains that the atom-light elementary frequency spectrum 
$\omega$, in terms of bare quantities, is given by the solutions of the equation
 \be
\det M(b) = 0
\label{eq:pas_renorm}
\ee
where the $3\times 3$ symmetric matrix $M$ is given by 
\begin{multline}
M_{\alpha\beta}(b) =(\omega_B^2-\omega^2)\delta_{\alpha\beta}\\
+\;\omega_{p,B}^2 \sum_{\KK\in\textrm{RL}} \frac{\delta_{\alpha\beta} K'^2 - K'_\alpha
K'_\beta}{(\omega/c)^2-K'^2}\;\cheta{\chi}^2(\KK';b),
\label{eq:Mfin}
\end{multline}
$\qq$ is the Bloch wavevector in the first Brillouin zone, 
$\KK$ is a vector of the reciprocal lattice (RL), 
$\KK'\equiv\KK-\qq$, 
\be
\omega_{p,B}^2=\frac{2\textrm{d}_B^2\omega_B}
{\hbar\varepsilon_0\mathcal{V}_{\textrm{L}}}, 
\ee
is the squared \emph{bare} plasmon frequency, 
$\mathcal{V}_{\textrm{L}}$ is the volume of the primitive unit cell 
of the lattice L in real space. 
We have taken the infinite quantization volume limit, so that
 $\mathcal{D}\to \mathbb{R}^3$, and 
the Bloch vector $\qq$ may assume any value in the first Brillouin zone.

Now, we specify the cut-off function $\cheta{\chi}(\kk;b)$ 
and renormalize the theory by expressing the equations in terms of physical realistic 
quantities $\omega_0$ and $d$, instead of the bare ones $\omega_B$ and $\textrm{d}_B$.
Inspired by \cite{Castin09}, we choose a Gaussian cut-off function \cite{dupdc,box}
\be
\cheta{\chi}^2(\kk;b)=e^{-k^2b^2}, \ \ b>0,
\label{eq:gauss}
\ee
which leads to a rapidly convergent sum in (\ref{eq:Mfin}).
If $b\rightarrow 0$, which is the limit we are interested in, the sum in (\ref{eq:Mfin}) diverges. 
The key idea is then that Eq.(\ref{eq:Mfin}) has been expressed in terms of \emph{bare} quantities. In what follows,
we renormalize the matrix $M_{\alpha\beta}$ by collecting the divergent terms and the bare quantities 
in new physical quantities.

\subsection*{Renormalization of the theory}

To this end, it is useful to introduce the quantity $\Delta_{\alpha\beta}(b)$ defined as: 
\begin{multline}
\Delta_{\alpha\beta}(b)=\delta_{\alpha\beta}\;\frac{\mathcal{V}_{\textrm{L}}(\omega /c)^3}{6\pi}\;\times\\
\left[\frac{1+2(\omega b/c)^2}{2\pi^{1/2}(\omega b/c)^3}-\frac{\textrm{Erfi}\,(\omega b/c)}{e^{(\omega b/c)^2}}\right]\\
+\sum_{\KK\in\textrm{RL}} \frac{\delta_{\alpha\beta} K'^2 - K'_\alpha
K'_\beta}{(\omega/c)^2-K'^2}\;e^{-K'^2b^2},
\label{eq:db}
\end{multline}
where $\KK'\equiv\KK-\qq$ and
$\textrm{Erfi}\,(x)=2\pi^{-1/2}\int_0^{x}dy\; \exp(y^2)$ is the imaginary error function. It has been shown in
\cite{Castin09} that $\Delta_{\alpha\beta}(b)$, which does not depend on bare quantities,
 has a finite limit $\Delta_{\alpha\beta}(0)$, as $b\rightarrow 0$ [see Eq.(\ref{eq:d0})]. 
It is then possible to write the matrix $M_{\alpha\beta}(b)$ in the limit of $b\rightarrow 0$ as:
\begin{multline}
M_{\alpha\beta}(b) =(\omega_B^2-\omega^2)\delta_{\alpha\beta}\\
+\frac{2 \textrm{d}_B^2 \omega_B}{\hbar\varepsilon_0}
 \left[\frac{\Delta_{\alpha\beta}(0)}{\mathcal{V}_{\textrm{L}}}-\frac{\delta_{\alpha\beta}}{12\pi^{3/2}b^3}-
\frac{\delta_{\alpha\beta}\;(\omega/c)^2}{6\pi^{3/2}b}\right]+O(b),
\label{eq:reg1}
\end{multline}
where we have used $\mathrm{Erfi}(\omega b/c)=O(b)$.
The key point is that,
although the last two terms in the square bracket $[\ldots]$ of (\ref{eq:reg1}) diverge as $b\rightarrow 0$, 
they diverge in a way not explicitly depending on the Bloch wavevector $\qq$
\cite{box}. In simple words,
considering the fact that the first term in the right-hand side of Eq.(\ref{eq:reg1}) 
is $(\omega_B^2-\omega^2)\delta_{\alpha\beta}$: The term inside the $[\ldots]$
diverging as $1/b^3$ may be summed to $\omega_B^2 \delta_{\alpha\beta}$, resulting in a renormalization
of $\omega_B$; the term inside the $[\ldots]$ diverging as $\omega^2/b$ may be summed to
$-\omega^2 \delta_{\alpha\beta}$, requiring also a renormalization of the atomic dipole moment $d_B$ 
appearing in the overall factor to the left of $[\ldots]$.
The fact that two parameters are renormalized in the Hamiltonian $H$ is reminiscent 
of the renormalization procedure of QED, where both the electron charge and mass
are renormalized \cite{Ryder}.

In practice, to apply the renormalization procedure,
we pull out in (\ref{eq:reg1}) the coefficient of the $-\omega^2$ term:
\be
M_{\alpha\beta}(b) = 
\left(1+\frac{\textrm{d}_B^2\omega_B}{3\pi^{3/2}\hbar\varepsilon_0c^2 b}\right)
\mathbb{M}_{\alpha\beta} + O(b),
\label{eq:form2}
\ee
where, by construction, the $3\times 3$ symmetric matrix $\mathbb{M}$ has a simple expression
in terms of the renormalized resonance frequency $\omega_0$ and the renormalized
dipole $\textrm{d}$:
\be
\mathbb{M}_{\alpha\beta} =(\omega_0^2-\omega^2)\delta_{\alpha\beta}\\
+\;\omega_{p}^2 \Delta_{\alpha\beta}(0), 
\label{eq:Mfinvera}
\ee
with the renormalized plasmon frequency
\be
\omega_{p}^2=\frac{2\textrm{d}^2\omega_0}{\hbar\varepsilon_0\mathcal{V}_{\rm L}},
\label{eq:opvera}
\ee
Identification of (\ref{eq:form2}) with (\ref{eq:reg1}) leads to the explicit
expressions for the two renormalized quantities \cite{conditions}
 \begin{eqnarray}
\label{eq:ren1}
 \omega_0^2 &=& \omega_B^2\;
 \left[1-\frac{\textrm{d}_B^2}{6\pi^{3/2}\hbar\varepsilon_0\omega_B b^3}\right]
 \left[1+\frac{\textrm{d}_B^2\omega_B}{3\pi^{3/2}\hbar\varepsilon_0c^2 b}\right]^{-1},\\
\textrm{d}^2 &=& \omega_0\; \frac{\textrm{d}_B^2}{\omega_B}
 \left[1-\frac{\textrm{d}_B^2}{6\pi^{3/2}\hbar\varepsilon_0\omega_B b^3}\right]^{-1}.
\label{eq:ren2}
 \end{eqnarray}
Finally, neglecting $O(b)$ in (\ref{eq:form2}),
we obtain that the atom-light elementary excitations for a generic Bravais lattice are the solution of the equation
\be
\det \mathbb{M} = 0
\label{eq:Mfinveraeq}
\ee
where $\mathbb{M}$ is given by (\ref{eq:Mfinvera}) and does not depend on the cut-off $b$.
In practical calculations of the matrix elements of $\mathbb{M}$, one uses the fact that
$\Delta_{\alpha\beta}(0)\equiv\lim_{b\rightarrow 0}\Delta_{\alpha\beta}(b)$ is related to $\Delta_{\alpha\beta}(b)$ of Eq.(\ref{eq:db})
by the useful expression \cite{Castin09}
\be
 \Delta_{\alpha\beta}(0)=\Delta_{\alpha\beta}(b)\;e^{(\omega b/c)^2},
 \label{eq:d0}
\ee
valid for all values of $b\ll a_{\rm min}$, where $a_{\rm min}$ is the minimum distance between two atoms.
In what follows, it will be convenient to use as a parameter, rather than the plasmon frequency,
the free space spontaneous emission rate of a single atom,
\be
\Gamma= \frac{d^2 k_0^3}{3\pi \varepsilon_0 \hbar}
\label{eq:Gamma}
\ee
where $k_0=\omega_0/c$ is the resonant wavevector.

\subsection*{Free-field solutions}

An important note is that, as already mentioned, the solutions of (\ref{eq:pas_renorm}) 
do not exhaust the spectrum.
One should add the so-called {\sl free-field solutions}
located on the free photon dispersion relations and 
corresponding to a free electric field of frequency $\omega$
(a solution of Maxwell's equations in the absence of matter)
that vanishes exactly in all the atomic positions. 
This was discussed in \cite{Knoester06} and also appears
in the calculations of Appendix \ref{app:exspect}.
In view of the discussion of spectral gaps to come, it is useful to keep in mind
the following lower bound on the frequency $\omega_{\rm free}$
of all possible free-field solutions \cite{demons}:
\be
\label{eq:borne}
\frac{\omega_{\rm free}}{c} \geq \inf_{\KK\in \textrm{RL}^*} \frac{K}{2}.
\ee
This inequality holds for an arbitrary (non necessarily Bravais) lattice,
and it is saturated for a Bravais lattice, as one can show using the two-mode
ansatz of \cite{Knoester06}.

\subsection*{Spectrum for a fcc lattice}

\begin{figure}[h]
\setlength{\unitlength}{.5in}
\begin{picture}(6,14)
\put(-0.2, 10.5){\includegraphics{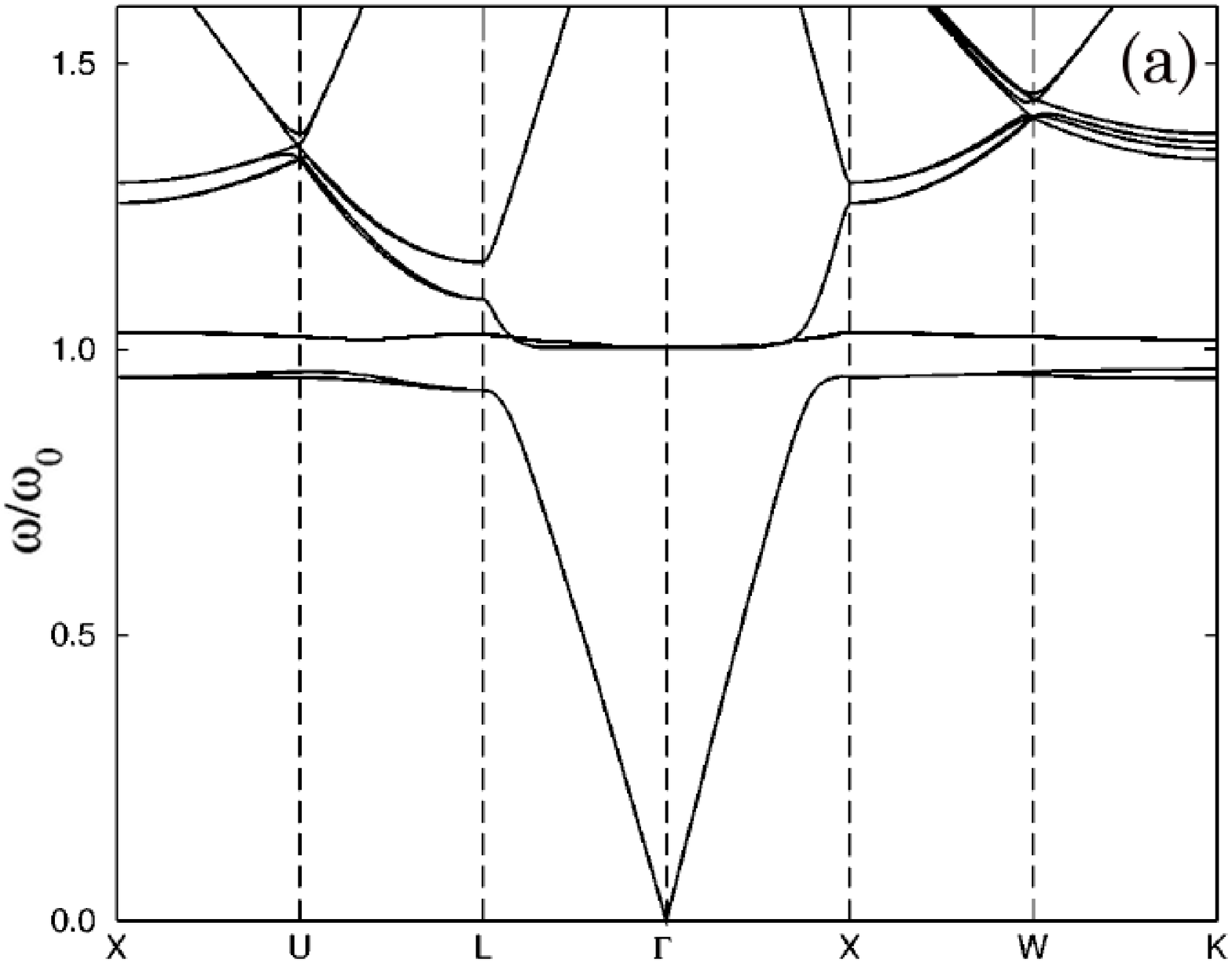}}
\put(-0.2, 5){\includegraphics{fig1b.eps}}
\put(-0.2, 0){\includegraphics{fig1c.eps}}
\end{picture}
\caption{Dispersion relation for light in a fcc atomic lattice,
with $k_0a\approx 5$, $\Gamma/\omega_0\approx 0.0167$, where $a$ is the lattice 
constant and $\Gamma$ the spontaneous emission rate
defined in (\ref{eq:Gamma}). In (a) the prediction found in figure (1) of \cite{Lagendijk96},  
in (b) the results of calculations obtained using Eq.(\ref{eq:Mfinveraeq}) of this paper, and 
(c) is a zoom of (b) around the atomic resonance frequency. 
(b) and (c) show the absence of an omnidirectional photonic gap, in contradiction with (a).
\label{fig:PRL96}
}
\end{figure}

\begin{figure}[h]
\setlength{\unitlength}{.5in}
\begin{picture}(6,14)
\put(-0.45, 5.0){\includegraphics{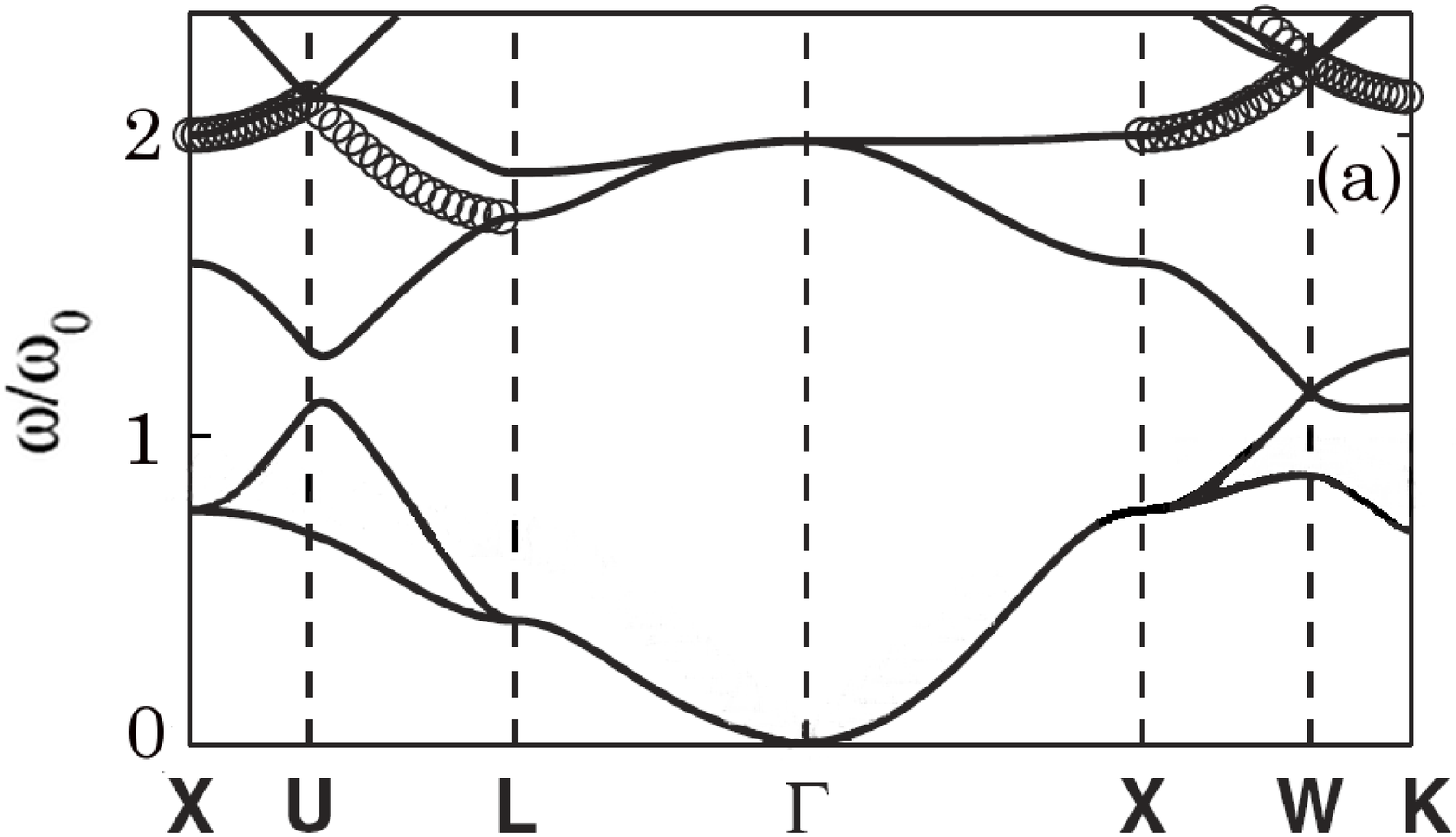}}
\put(-0.2, 0){\includegraphics{fig2b.eps}}
\end{picture}
\caption{(Color online) Dispersion relation for light in a fcc atomic lattice,
with $k_0a\approx 3.14$, 
$\Gamma/\omega_0\approx 1.189$, where $a$ is the lattice constant. (a) is the prediction found in 
figure (2) of \cite{Knoester06} [circles correspond to the free modes, i.e. modes of the free field
that vanish at all atomic positions],  
(b) is the result of calculations using Eq.(\ref{eq:Mfinveraeq}) of this paper 
[the red-dashed lines correspond to the free modes]. (a) and (b) appear to be identical.
\label{fig:PRL06}
}
\end{figure}

As a straightforward application of our approach, we calculate the light dispersion
relation in the fcc lattice.
There are two existing predictions, \cite{Lagendijk96} and \cite{Knoester06},
that reach opposite conclusions concerning the existence of an omnidirectional gap.
We have calculated the dispersion relation for exactly the same parameters
as these two references, see Fig.\ref{fig:PRL96} for \cite{Lagendijk96}
and Fig.\ref{fig:PRL06} for \cite{Knoester06} respectively, which allows a direct comparison.
Our theory disagrees with the result of \cite{Lagendijk96}; it quantitatively agrees 
with the one of \cite{Knoester06} \emph{provided} one replaces in the final result of \cite{Knoester06}
(but of course not in the Lagrangian of \cite{Knoester06})
the bare values of the atomic and plasmon frequencies by their renormalized values.
We conclude, as in \cite{Knoester06}, that there is no spectral gap for light in a fcc atomic lattice.

\section{Light dispersion relation in arbitrary atomic crystals}
\label{sec:nonBravais}

Let us consider the generic case of a crystal which is not a Bravais lattice, 
that is it has more than one atom per primitive unit cell;
it can be seen as the composition of $P$  translated copies of the same Bravais lattice. 
The crystal is then the periodic repetition of an elementary base of $P$ atoms. 
In this case it is possible to generalize the equations obtained for a Bravais lattice,
as done in the appendix, and to obtain the equation
\be
\det \mathbb{M}= 0
\label{eq:Mfinveramultieq}
\ee
where the $3P\times 3P$ hermitian matrix $\mathbb{M}$ is given by
\be
\mathbb{M}_{\alpha\mu,\beta\nu} =(\omega_0^2-\omega^2)\delta_{\alpha\beta}\delta_{\mu\nu}\\
+\;\omega_{p}^2 \Delta_{\alpha\mu,\beta\nu}(0)
\label{eq:Mfinveramulti}
\ee
with $\alpha,\beta\in\{x,y,z\}$, $\mu,\nu\in\{1,\cdots,P\}$,  
and $\omega_{p}^2$ is given by Eq.(\ref{eq:opvera}). 
We have defined
%\begin{widetext}
\begin{multline}
\Delta_{\alpha\mu,\beta\nu}(b)=\delta_{\alpha\beta}\delta_{\mu\nu}\;\frac{\mathcal{V}_{\textrm{L}}(\omega /c)^3}{6\pi}\;\times\\
\;\left[\frac{1+2(\omega b/c)^2}{2\pi^{1/2}(\omega b/c)^3}-\frac{\textrm{Erfi}(\omega b/c)}{e^{(\omega b/c)^2}}\right]\\
+\sum_{\KK\in\textrm{RL}}e^{-i\KK'\cdot(\rr_\mu-\rr_\nu)}\; \frac{\delta_{\alpha\beta}\; K'^2 - K'_\alpha
K'_\beta}{(\omega/c)^2-K'^2}\;e^{-K'^2b^2},
\label{eq:deltamulti}
\end{multline}
%\end{widetext}
where $\KK'\equiv\KK-\qq$ and the sum is over the reciprocal lattice
of the underlying Bravais lattice.
The index $\mu$ or $\nu$ labels the $P$ atoms inside the primitive unit cell,
and $\rr_\mu$ and $\rr_\nu$ are the positions of the corresponding
atoms in that cell. 
By a slight generalization of the technique developed in \cite{Castin09}, 
it is possible to show that
\be
 \Delta_{\alpha\mu,\beta\nu}(0)=
\Delta_{\alpha\mu,\beta\nu}(b)\;e^{(\omega b/c)^2},\;\;\;\;\;\;\;\;\;\;\;\; b\ll a_{\rm min},
 \label{eq:domulti}
\ee
where $a_{\rm min}$ is the minimal distance between two atoms in the crystal.
In particular, this implies that,  for $\mu\neq\nu$, the sum over $\KK$ in
(\ref{eq:deltamulti}) has a finite limit when $b\to 0$.

Here again, we note that {\sl free-field} solutions, that is eigenmodes of the free field
with an electric field that vanishes in all atomic locations, may have been left
out from (\ref{eq:Mfinveramultieq}) and should be investigated separately.
The lower bound (\ref{eq:borne}) on the possible free-field eigenfrequencies
$\omega_{\rm free}$ still applies \cite{demons}, but, in the non-Bravais lattice case, this lower
bound is in general not reached, since the vanishing of the electric field 
on each atomic location $\rr_\mu$ inside the primitive unit cell adds extra constraints
with respect to the Bravais case.

\section{The long wavelength limit: generalized Lorentz-Lorenz equation}
\label{sec:lorentz}

Let us consider the long wavelength (LW) limit of the energy spectrum, for a general (even non-Bravais) lattice.
We expand Eq.(\ref{eq:Mfinveramulti}) in the limit $qa\ll1$ and $\omega a/c\ll1$,
where $a$ is the minimal distance between two atoms in the underlying Bravais lattice. 
The resulting equation is
\be
\det \mathbb{M}^{\textrm{LW}}=0,
\label{eq:LW}
\ee
where the matrix $\mathbb{M}^{\textrm{LW}}$ is given by
\begin{multline}
\mathbb{M}_{\alpha\mu,\beta\nu}^{\textrm{LW}} =(\omega_0^2-\omega^2)\delta_{\alpha\beta}\delta_{\mu\nu}\\
+\omega_{p}^2 \;\frac{\delta_{\alpha\beta}\; q^2 - q_\alpha
q_\beta}{(\omega/c)^2-q^2} +\omega_{p}^2\;\textrm{J}_{\alpha\mu,\beta\nu},
\label{eq:MfinveramultiLW}
\end{multline}
with
\begin{multline}
\textrm{J}_{\alpha\mu,\beta\nu}=\lim_{b\rightarrow 0}\left[ \delta_{\alpha\beta}\delta_{\mu\nu}\;\frac{\mathcal{V}_{\textrm{L}}}{12\pi^{3/2}\;b^3}\right.\\
\left.-\sum_{\KK\in\textrm{RL}^*}e^{-i\KK\cdot(\rr_\mu-\rr_\nu)}\; \frac{\delta_{\alpha\beta}\; K^2 - K_\alpha
K_\beta}{K^2}\;e^{-K^2b^2}\right].
\end{multline}
The matrix $\delta_{\alpha\beta} q^2 - q_\alpha
q_\beta$ in Eq.(\ref{eq:MfinveramultiLW}) is proportional to a projector:
it has eigenvalues equals to zero or to $q^2$, respectively for the modes 
which have a longitudinal or transverse polarization with respect to $\qq$. 
Equation (\ref{eq:LW}) is a generalization of the result obtained by
Hopfield in Eq.(30) of \cite{Hopfield58},  where the excitation spectrum 
in the long wavelength limit was calculated for a cubic crystal.
Since the cubic crystal is a Bravais lattice, we can drop the indices $\mu$ and $\nu$;  
using the symmetries of the cubic structure one finds that  
$\textrm{J}_{\alpha,\beta}=2\delta_{\alpha\beta}/3$. With this value 
one obtains that Eq.(\ref{eq:LW}) gives both the longitudinal mode frequency ignored in
\cite{Hopfield58},
\be
\omega^{\textrm{LW,cubic}}_{\parallel}=\sqrt{\omega_0^2 +\frac{2}{3} \omega_p^2},
\ee
and the transverse mode frequencies, solution as in \cite{Hopfield58}
of the Lorentz-Lorenz (or Clausius-Mossotti) equation 
 \be
\frac{q^2}{(\omega^{\textrm{LW,cubic}}_{\perp}/c)^2}=1+\frac{n\alpha(\omega^{\textrm{LW,cubic}}_{\perp})}{1-\frac{1}{3}\;n\alpha(\omega^{\textrm{LW,cubic}}_{\perp})}
\label{eq:HT}
\ee
where $n=1/\mathcal{V}_{\rm L}$  
is the atomic density of the cubic lattice and the \emph{real} function $\alpha(\omega)$ is an atomic polarizability
\be
n\alpha(\omega)=\frac{\omega_{p}^2}{\omega_0^2-\omega^2}.
\label{eq:alpha}
\ee
Then Eq.(\ref{eq:LW}) is a generalized Lorentz-Lorenz equation to any periodic structure.

\begin{figure}[tbh]
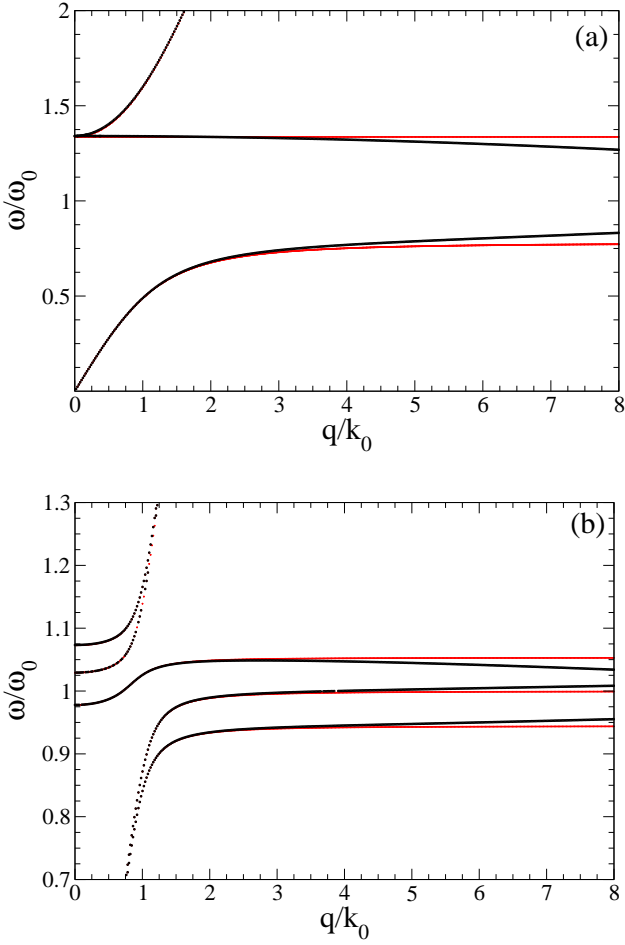

\begin{center}
\includegraphics[width=0.50\textwidth,clip=]{fig3a.eps}
\includegraphics[width=0.50\textwidth,clip=]{fig3b.eps}
\end{center}
\caption{(Color online) Dispersion relation for light, for
the approximate long wavelength expression of Eq.(\ref{eq:LW}) (red), and for 
the complete expression Eq.(\ref{eq:Mfinveraeq}) (black). The spectrum is calculated as a 
function of $q/k_0$ along direction $(1,1,1)$ in reciprocal space.
In (a) the simple cubic lattice with $k_0a=1/5$ and $\Gamma/\omega_0=5\times 10^{-4}$, where $a$ 
is the lattice constant. 
The long  wavelength approximation fails when the horizontal (degenerate) branches of the exact spectrum 
start approaching each other, eventually closing the gap at larger values of $q/k_0$. 
In (b) an orthorhombic lattice with basis vectors in the direct space along $x,y,z$ 
with norms respectively proportional to $2\pi$, $\pi$, and $2^{3/2}\pi$; here $k_0a=1/5$ 
and $\Gamma/\omega_0=5\times 10^{-5}$, where $a=\mathcal{V}_{\textrm{L}}^{1/3}$ and
 $\mathcal{V}_{\textrm{L}}$ is the volume of the primitive unit cell.  
Five non-degenerate branches are present, and the long  wavelength approximation 
fails when the three horizontal branches of the exact spectrum start approaching each other, 
eventually closing the gap at larger values of $q/k_0$.
\label{fig:hopfield}}
\end{figure}

To illustrate these results, we have plotted in Fig.\ref{fig:hopfield}
the dispersion relation for light obtained in the long-wavelength approximation
(\ref{eq:LW})
and from the exact equation (\ref{eq:Mfinveramultieq}), for a cubic lattice
and for an orthorhombic lattice.
As expected, the long wavelength approximation becomes inaccurate
for $1\lesssim qa$.

As a side remark, it is worth stressing that one can relate the tensor $\textrm{J}$ to the dipolar electrostatic
energy of the lattice. For a Bravais lattice $L$ one obtains \cite{derivation}
\be
\frac{1}{\mathcal{V}_L} \left[\textrm{J}_{\alpha\beta} - \frac{2}{3}\delta_{\alpha\beta}\right] =
\sum_{\RR\in L^*} \frac{\delta_{\alpha\beta}-3R_\alpha R_\beta/R^2}{4\pi R^3}.
\label{eq:electro}
\ee
The sum in the right-hand side of this equation is not absolutely convergent, so that its precise meaning
needs to be defined, as we did in \cite{derivation}.

\section{The diamond crystal: omnidirectional photonic band-gap}
\label{sec:diamant}

As pointed out in the introduction, no atomic Bravais lattice leading
to an omnidirectional gap for light is known. A natural idea to obtain such
a gap is thus to consider lattices with several atoms per primitive unit cell.
If one replaces the point-like atoms with extended objects like dielectric spheres,
it is known that the diamond lattice leads to a photonic gap provided that the spheres
fill a large enough fraction of the unit cell \cite{Soukoulis90}.

We have therefore calculated the band structure for light in an diamond atomic lattice,
solving numerically (\ref{eq:Mfinveramultieq}) 
of section \ref{sec:nonBravais}. The diamond lattice is formed
by the superposition of two copies of the same Bravais lattice: the fcc lattice
of lattice constant $a$, generated by the three basis vectors
$\mathbf{e}_1=(0,a/2,a/2)$, $\mathbf{e}_2=(a/2,0,a/2)$, $\mathbf{e}_3=(a/2,a/2,0)$,
 and a second fcc lattice obtained by translating the first lattice by the vector $(a/4,a/4,a/4)$.
For the reciprocal lattice of the fcc lattice, we take the basis
$\tilde{\mathbf{e}}_1=(-2\pi/a,2\pi/a,2\pi/a)$, $\tilde{\mathbf{e}}_2=(2\pi/a,-2\pi/a,2\pi/a)$, 
$\tilde{\mathbf{e}}_3=(2\pi/a,2\pi/a,-2\pi/a)$.
In view of application to atomic gases, we restrict to the perturbative limit $\omega^2_p/\omega_0^2\to 0$,
around the atomic resonance frequency $\omega_0$,
calculating $\omega-\omega_0$ to first order in $\omega_p^2$. In this regime,
it is convenient to take as a unit of frequency the spontaneous emission
rate $\Gamma$ of a single atom, defined in (\ref{eq:Gamma}),
and one finds that $\omega-\omega_0$ is any of the eigenvalues
of the matrix  ${\mathbb P}_{\alpha\mu,\beta\nu}(0) \equiv \lim_{b\to 0}
{\mathbb P}_{\alpha\mu,\beta\nu}(b)$, with 
\begin{multline}
\label{eq:cas_pertu}
{\mathbb P}_{\alpha\mu,\beta\nu}(b) =  \frac{\Gamma}{2} \delta_{\alpha\beta} \delta_{\mu\nu}
\left[\frac{1+2(k_0b)^2}{2\pi^{1/2} (k_0b)^3}-\textrm{Erfi}\,(k_0b) e^{-k_0^2b^2}\right]
\\
+ \frac{3\pi\Gamma}{k_0^3\mathcal{V}_\textrm{L}}
\sum_{K\in \textrm{RL}} e^{-i\KK'\cdot(\rr_\mu-\rr_\nu)}
\frac{\delta_{\alpha\beta} K'^2 -K'_\alpha K'_\beta}{k_0^2-K'^2}
\, e^{-K'^2 b^2}
,
\end{multline}
where $\KK'\equiv\KK-\qq$, $k_0=\omega_0/c$ is the resonant wavevector,
and where the limit for $b\to 0$ is computed in practice
from the relation ${\mathbb P}_{\alpha\mu,\beta\nu}(0)
={\mathbb P}_{\alpha\mu,\beta\nu}(b)\, e^{k_0^2 b^2}$ for $b\ll a_{\rm min}=a\sqrt{3}/4$
($b=0.05a$ in the present calculations).
Clearly, in this perturbative limit,
$(\omega-\omega_0)/\Gamma$ only depends on the dimensionless quantities
$\qq a$ and $k_0 a$.

\begin{figure}[htb]
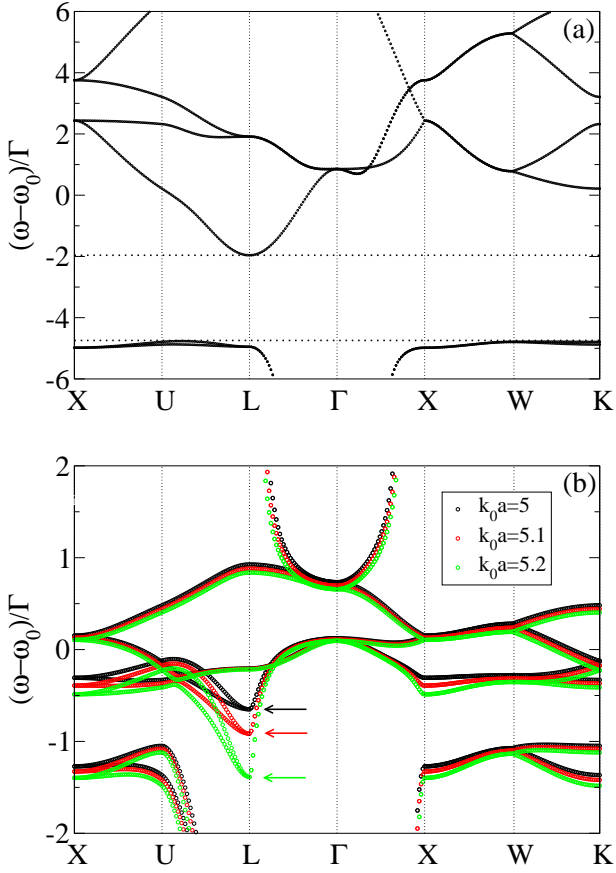

\includegraphics[width=0.45\textwidth,clip=]{fig4a.eps}
\includegraphics[width=0.45\textwidth,clip=]{fig4b.eps}
\caption{(Color online) Dispersion relation for light in a diamond atomic lattice as a function
of the Bloch vector along the standard irreducible path in the first Brillouin zone,
in the perturbative regime $\omega_p^2/\omega_0^2 \ll 1$ (see text).
(a) $k_0 a=2$, with $k_0=\omega_0/c$ and $a$ is the diamond lattice constant
(see text). Symbols: numerical values of the allowed frequencies $\omega$.
Dashed horizontal lines: lower and upper borders of the gap obtained from the
full density of states (not restricting to the irreducible path).
(b) For several values of $k_0 a$ close to the vanishing of the gap: $k_0 a=5$
(black symbols), $k_0 a=5.1$ (red symbols), $k_0 a=5.2$ (green symbols).
\label{fig:pir}
}
\end{figure}

\begin{figure}[htb]
\includegraphics[width=0.45\textwidth,clip=]{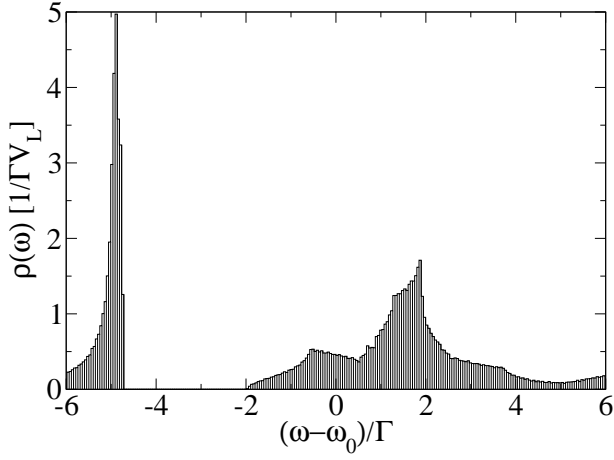}
\caption{Density of states $\rho(\omega)$ for light in an diamond atomic lattice,
in the perturbative regime $\omega_p^2/\omega_0^2 \ll 1$,
see (\ref{eq:rho}).
The whole unit cell of the reciprocal lattice is explored to obtain
this density of states, with a mesh of $10^6$ points.
The histogram (with 250 bins) clearly reveals the existence
of an omnidirectional gap at frequencies below the atomic resonance frequency.
We have taken  $k_0 a=2$ as in Fig.\ref{fig:pir}a.
$\rho(\omega)$ is in units of $1/(\Gamma \mathcal{V}_L)$
where $\Gamma$ is the single atom spontaneous emission rate
and $\mathcal{V}_L=a^3/4$ is the volume of the unit cell of the fcc lattice.
\label{fig:dos}
}
\end{figure}

We investigated first the arbitrarily chosen value $k_0 a=2$. In Fig.\ref{fig:pir}a, we give
$\omega-\omega_0$ as a function of $\qq$ along the standard irreducible path in the first Brillouin
zone of the fcc lattice: a gap appears. To prove the existence of a truly omnidirectional gap,
exploring the irreducible path is not sufficient. We thus evaluated the full density of
states $\rho(\omega)$, 
\be
\rho(\omega) = \sum_{n=1}^{6} \int_{\rm FBZ} \frac{d^3q}{(2\pi)^3} \, 
\delta(\omega-\omega_{\qq,n})
\label{eq:rho}
\ee
where the integral over $\qq$ is taken in the first Brillouin
zone, the sum runs over the band index $n$, $\omega_{\qq,n}$ is the light dispersion
relation in band $n$.
Numerically, we have explored the whole unit cell of the reciprocal lattice,
parametrized as $\qq=\sum_{\alpha} Q_\alpha \tilde{\mathbf{e}}_\alpha$,
$\QQ\in [-1/2,1/2)^3$,  with a cubic mesh for $\QQ$ with $10^6$ points.
The result in Fig.\ref{fig:dos} unambiguously shows the presence of an omnidirectional gap,
$\rho(\omega)$ vanishing over an interval $[\omega_{\rm inf},\omega_{\rm sup}]$.
The values of $\omega_{\rm inf}$ and $\omega_{\rm sup}$ are reported as horizontal
dashed lines in Fig.\ref{fig:pir}a, showing that remarkably the partial gap on the irreducible
path is very close to the full omnidirectional gap.

In a second stage, we have varied the lattice spacing $a$. The variation of the gap
$\Delta\equiv \omega_{\rm sup}-\omega_{\rm inf}$ 
as a function of $k_0 a$ is given in Fig.\ref{fig:gap}a, revealing that there
exists no gap for too high values of $k_0 a$, $5.14 \lesssim k_0 a$.
This may be understood by looking again at the light dispersion relation along the irreducible
path of the first Brillouin zone, for values of $k_0 a$ around 5.1, see Fig.\ref{fig:pir}b:
Around point $L$, defined by $q_L=(\pi/a,\pi/a,\pi/a)$, a frequency branch is seen to rapidly move towards low frequencies
when $a$ increases, leading to a closure of the gap.
The modulus of the Bloch vector is $q_L=\pi\sqrt{3}/a\simeq 5.44/a$, and when $q_L$ approaches
the value $k_0$ from above for increasing $a$, the denominator 
$k_0^2-q_L^2$ appearing in (\ref{eq:cas_pertu}) for $\KK=\mathbf{0}$
becomes very small and negative, 
leading to a mode frequency $\omega$ close to the $L$ point more and more
below the atomic resonance frequency $\omega_0$.
In the opposite limit of a small lattice spacing $a$, that is of an increasing atomic density,
one finds that the gap increases; from the overall factor $1/(k_0 a^3)$ appearing in 
(\ref{eq:cas_pertu}), one naively expects a gap scaling as $1/a^3$ in this limit, which
is indeed roughly the case.

The previous discussion can be extended to the variation of the gap boundaries
$\omega_{\rm inf}$ and $\omega_{\rm sup}$, plotted as functions of $k_0 a$ in
Fig.\ref{fig:gap}b. The log-log scale reveals that $\omega_{\rm inf}$ and $\omega_{\rm sup}$
approximately vary as power laws with $a$ at low $a$, and the corresponding slopes
$\sim -3.25$ and $\sim -2.8$ (see the dotted lines)
indeed indicate exponents close to $-3$.
At large $a$, the rapid variation of $\omega_{\rm sup}$ leading to the gap closure
around $\omega_{\rm inf}-\omega_0= \omega_{\rm sup}-\omega_0 \simeq -\Gamma$
is also quite apparent.

To be complete, let us mention that the previous discussion about the existence
of a spectral gap is not affected by the free-field solutions.
From (\ref{eq:borne}) we indeed find that any free-field solution 
has a frequency $\omega_{\rm free}\geq c\, \inf_{\KK\in \textrm{RL}^*} K/2=
c|\tilde{\mathbf{e}}_1+\tilde{\mathbf{e}}_2+\tilde{\mathbf{e}}_3|/2=
\pi \sqrt{3}c/a$. We have seen that having a spectral gap
requires $k_0 < q_L  = \pi\sqrt{3}/a$, which thus implies $\omega_0 < \omega_{\rm free}$
\cite{question}.

\subsection*{Experimental realizability of a diamond atomic lattice}

To experimentally test the presence of a gap for light in a diamond
lattice with real atoms, the first step is to use an atomic species with
a transition between a ground state with a spin $J_g=0$ and an excited
state with a spin $J_e=1$, so that the harmonic oscillator representation
of the atomic dipole in the Hamiltonian $H$ is reasonable.
A natural candidate is strontium ${}^{88}$Sr, already used in experiments
on light coherent backscattering \cite{David02}, but there are
of course other possibilities, such as bosonic ytterbium ${}^{174}$Yb
where a Bose-Einstein condensate is available \cite{Yabu03}.
The second step is to realize a diamond structure, by loading the atoms
in an appropriate optical lattice with a filling factor one (Mott
phase).
Very recently, such a Mott phase with ${}^{174}$Yb was realized
in a cubic lattice \cite{Takahashi09}.

\begin{figure}[htb]
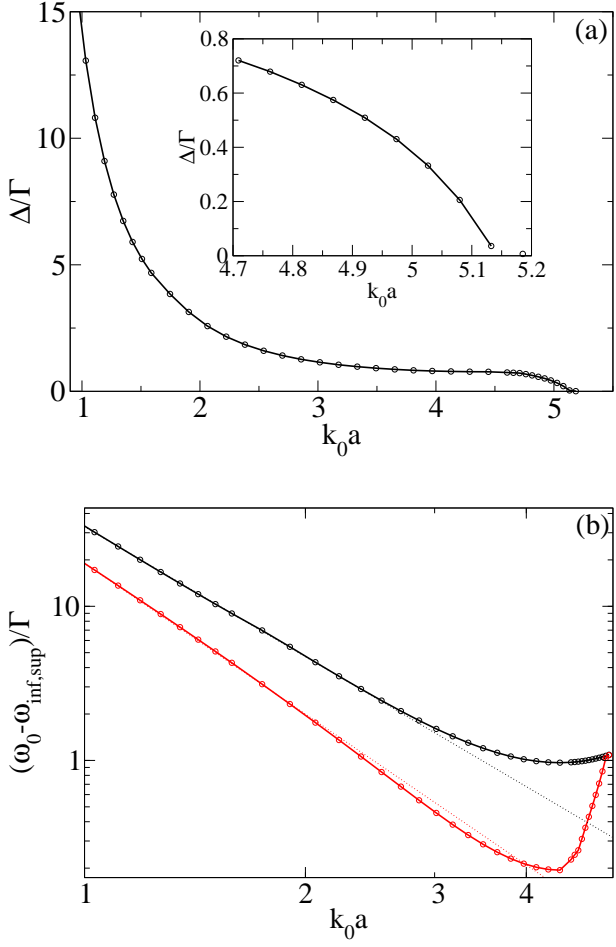

\includegraphics[width=0.45\textwidth,clip=]{fig6a.eps}
\includegraphics[width=0.45\textwidth,clip=]{fig6b.eps}
\caption{(Color online) For the light spectrum in a diamond atomic lattice,
in the perturbative limit $\omega_p^2\ll \omega_0^2$, and as a function of $k_0 a$,
(a) value of the omnidirectional
spectral gap $\Delta=\omega_{\rm sup}-\omega_{\rm inf}$, 
and (b) positions of the lower frequency $\omega_{\rm inf}$
and upper frequency $\omega_{\rm sup}$ of the gap with respect to the atomic resonance frequency.
$a$ is the lattice constant (see text), $k_0=\omega_0/c$ and $\Gamma$ is the spontaneous
emission rate defined in (\ref{eq:Gamma}).
The inset in (a) is a magnification.
\label{fig:gap}
}
\end{figure}

Atomic lattices with two atoms per primitive unit cell have already
been realized experimentally \cite{Phillips07}, however to our knowledge
not with a diamond structure.
This missing element, the diamond lattice, can be realized
building on the ideas of \cite{Grynberg94}, as was shown
in \cite{John04} in the different context of holographic
lithography of dielectric materials \cite{Courtois99}.
We recall here briefly the idea: The optical lattice is produced
by the coherent superposition of four laser plane waves,
resulting in the total laser field of positive frequency part
\be
\mathbf{E}^{(+)}_{\rm laser}(\rr,t) = \sum_{j=0}^{3} \mathbf{E}_j e^{i(\kk_j\cdot\rr
-\omega_{\rm laser}t)},
\ee
where the laser frequency $\omega_{\rm laser}$ is very far from 
the atomic resonance frequency $\omega_0$ so that the optical
lattice is a purely conservative potential.
The vectorial amplitudes of the laser plane waves have to satisfy
the transversality conditions
\be
\label{eq:cond0}
\kk_j \cdot \mathbf{E}_j =0 \ \forall j.
\ee
The lightshift experienced by the atoms in this laser field
is proportional to the laser intensity $I_{\rm laser}(\rr)
\propto |\mathbf{E}^{(+)}_{\rm laser}|^2$.
Taking for simplicity linearly polarized plane waves, the vectorial amplitudes
$\mathbf{E}_j$ may be chosen real by a convenient choice of the origin
of coordinates \cite{Grynberg94}, so that
\be
|\mathbf{E}^{(+)}_{\rm laser}|^2
=
\left(\sum_{j=0}^{3} E_j^2\right)
+\sum_{0\leq j\leq j'\leq 3}  \;
2\mathbf{E}_j \cdot \mathbf{E}_{j'}
\cos[(\kk_j-\kk_{j'})\cdot\rr].
\ee
The laser wavevectors $\kk_j$ are chosen so that the $\kk_j-\kk_{j'}$
generates the reciprocal lattice of the fcc lattice. As in \cite{John04} one may choose
$\kk_0=\frac{\pi}{a}(0,-2,-1)$,
$\kk_1=\frac{\pi}{a}(2,0,1)$,
$\kk_2=\frac{\pi}{a}(0,2,-1)$, and
$\kk_3=\frac{\pi}{a}(-2,0,1)$. These four vectors have the same
modulus, as it should be, which relates the value of the laser
frequency to the diamond lattice constant,
\be
k_{\rm laser} = \frac{\omega_{\rm laser}}{c} = \frac{\sqrt{5}\pi}{a}.
\ee
To have a significant spectral gap, see Fig.\ref{fig:gap},
one should satisfy the condition
$k_0 a < 5$, that is $\omega_{\rm laser} > 1.4 \omega_0$, which corresponds to
a blue detuned lattice, where the atoms are trapped in
the minima of the laser intensity.

The challenge is now to correctly choose the vectorial amplitudes
$\mathbf{E}_j$ to ensure that the minima of $I_{\rm laser}(\rr)$ 
form a diamond lattice. An elegant solution was given in \cite{John04}:
by imposing the conditions 
\bea
\label{eq:cond1}
\mathbf{E}_3\cdot \mathbf{E}_2 &=& \mathbf{E}_3\cdot \mathbf{E}_0 = 
\mathbf{E}_2 \cdot \mathbf{E}_1 = - \mathbf{E}_1\cdot \mathbf{E}_0 >0,
\\
\mathbf{E}_3\cdot \mathbf{E}_1 &=& \mathbf{E}_2 \cdot \mathbf{E}_0 =0,
\label{eq:cond2}
\eea
one realizes an intensity pattern
\be
I_{\rm laser}(\rr) = I_0 + I_1 \left[
-\cos(\sum_{\gamma=1}^{3}\tilde{\mathbf{e}}_\gamma\cdot \rr)
+\sum_{\gamma=1}^{3} \cos(\tilde{\mathbf{e}}_\gamma\cdot\rr)\right]
\label{eq:ip}
\ee
where  $I_0>0$, $I_1>0$,  
and the basis vectors $\mathbf{\tilde{e}}_\gamma$ of the reciprocal lattice
are given in the beginning of this section \ref{sec:diamant}.
The solutions of (\ref{eq:cond0}),(\ref{eq:cond1}),(\ref{eq:cond2}) are not unique.
If one fixes the value of $I_1$ (that is 
the value of $\mathbf{E}_3\cdot \mathbf{E}_2$)
to suppress a global scaling invariance, we are left with 10 equations
for the 12 unknown components of the $\mathbf{E}_j$'s, leading
to an actually explicitly calculable continuum of solutions parametrized
by two real parameters. One of these continuous parameters
corresponds to the invariance of all the equations 
by the scaling transform $\mathbf{E}_j\to \mathbf{E}_j/\lambda$,
$j\in\{1,3\}$, and $\mathbf{E}_j\to \mathbf{E}_j\lambda$,
$j\in\{0,2\}$, with the scaling factor $\lambda \in \mathbb{R}^*$.
The particular solution given in \cite{John04} obeys
the two constraints $E_1^2=E_2^2=E_3^2$ presumably added for experimental 
convenience.

The key point is then that the
minima of the intensity pattern (\ref{eq:ip}) are 
located in $(3a/8,3a/8,3a/8)$ and in $(5a/8,5a/8,5a/8)$ modulo
any vector of the fcc lattice \cite{maxima}. 
These minima correspond to the same
laser intensity, and lead to local harmonic microtraps that are
isotropic. Since the relative vectorial position of these
two minima is $(a/4,a/4,a/4)$,
the set of all laser intensity minima indeed form a diamond lattice.

\section{Conclusion}
\label{sec:conclusion}

We have investigated the Fano-Hopfield model for propagation of light in 
a periodic ensemble
of atoms with fixed positions 
for the most general (even non-Bravais) lattice geometry and without restricting to the long
wavelength regime for light. We have shown that all divergences in the large momentum
cut-off limit may be eliminated in a systematic way by a renormalization of two parameters of the 
Hamiltonian, the atomic resonance frequency and the atomic dipole. 

First, we have used our theory to quantitatively confirm the predictions of \cite{Knoester06}
for the spectrum of light in a face-centered cubic lattice and the absence of an omnidirectional
spectral gap in that case.
Second, we have shown that a diamond atomic lattice can lead to an omnidirectional gap.
Since a diamond atomic lattice may be realized in practice, as we discussed,
this opens up the possibility of observing such a spectral gap for light 
in an ultracold atomic ensemble, which would bridge the state of the art 
gap between atomic gases and photonic band-gap solid state materials.

\begin{acknowledgments}
We thank Iacopo Carusotto and Christophe Salomon for useful discussions and suggestions.
One of us (M.A.) thanks ANR Gascor and IFRAF for financial support. 
Our group at Laboratoire Kastler Brossel is a member of IFRAF.

\end{acknowledgments}

\appendix

\section{Derivation of the excitation spectrum in terms of bare quantities \label{app:exspect}}

\subsection{Equations of motion}
\label{subsec:eqom}

Starting from $H$ in (\ref{eq:H}), and from the bosonic commutation relations 
it is possible to derive the system of coupled equations of motion in the 
Heisenberg picture for the matter excitation operators 
$\hat{b}_{i,\alpha}$ ($1\leq i \leq N$ and $\alpha\in\{x,y,z\}$),
 and for the light excitation operators $\hat{a}_{\kk,\epsbs}$
($\kk\in\mathcal{D}$ and $\epsb\perp\kk$):
\begin{eqnarray}
i\hbar\,\frac{d}{dt}\, \hat{b}_{i,\alpha}&=&\hbar\omega_B\,\hat{b}_{i,\alpha}-\textrm{d}_B\,
\hat{{\mathbb E}}_{\perp,\alpha}(\RR_i),\label{eq:em1}\\
i\hbar\,\frac{d}{dt}\, \hat{b}^{\dagger}_{i,\alpha}&=&-\hbar\omega_B\,
\hat{b}^{\dagger}_{i,\alpha}+\textrm{d}_B\,\hat{{\mathbb E}}_{\perp,\alpha}(\RR_i),\label{eq:em2}\\
i\hbar\,\frac{d}{dt}\, \hat{a}_{\kk\epsbs} &=& \hbar ck\, \hat{a}_{\kk\epsbs}\notag\\
&&\!\!\!\!\!\!\!\!\!\!\!\!\!\!\!\!\!\!\!\!\!\!\!\!-\;\mathcal{C}_k\;\cheta{\chi}(\kk;b) \sum_{i=1}^{N} \sum_{\alpha}^{x,y,z} (\hat{b}_{i,\alpha}+ 
\hat{b}^{\dagger}_{i,\alpha})\,\eps_{\alpha}^*\, e^{-i\kk\cdot \RR_i},\label{eq:em3}\\
i\hbar\,\frac{d}{dt}\, \hat{a}^{\dagger}_{\kk\epsbs} &=& -\hbar ck\, 
\hat{a}^{\dagger}_{\kk\epsbs}\notag\\
&&\!\!\!\!\!\!\!\!\!\!\!\!\!\!\!\!\!\!\!\!\!\!\!\!-\;\mathcal{C}_k\;\cheta{\chi}(\kk;b) \sum_{i=1}^{N} \sum_{\alpha}^{x,y,z} (\hat{b}_{i,\alpha}
+ \hat{b}^{\dagger}_{i,\alpha})\,\eps_{\alpha}\, e^{i\kk\cdot \RR_i}\label{eq:em4},
\end{eqnarray}
where $\mathcal{C}_{k}=-\textrm{d}_B\,\mathcal{E}_k$ and
$\mathcal{E}_k=i [\hbar kc/(2\varepsilon_0\textrm{V})]^{1/2}$.
The standard Bogoliubov procedure to obtain the eigenmodes amounts to replacing the operators
in the equations of motion with complex numbers oscillating in time at the
eigenfrequency $\omega$:
\begin{eqnarray}
\hat{b}_{i,\alpha} &\rightarrow& U_{i,\alpha}\;\;e^{-i\omega t},\\
\hat{b}^{\dagger}_{i,\alpha} &\rightarrow& V_{i,\alpha}\;\;e^{-i\omega t},\\
\hat{a}_{\kk\epsbs} &\rightarrow& u_{\kk\epsbs}\;e^{-i\omega t},\\
\hat{a}^{\dagger}_{\kk\epsbs} &\rightarrow& v_{\kk\epsbs}\;e^{-i\omega t}.
\end{eqnarray}

The set of equations (\ref{eq:em1})-(\ref{eq:em4}) then becomes
\begin{eqnarray}
\hbar\omega\, U_{i,\alpha}&=&\hbar\omega_B\,U_{i,\alpha}-\textrm{d}_B\,
{\mathbb E}_{\perp,\alpha}(\RR_i),\label{eq:em12}\\
\hbar\omega\, V_{i,\alpha}&=&-\hbar\omega_B\,V_{i,\alpha}+\textrm{d}_B
\,{\mathbb E}_{\perp,\alpha}(\RR_i),\label{eq:em22}\\
\hbar\omega\, u_{\kk\epsbs} &=& \hbar ck\, u_{\kk\epsbs}\notag\\
&&\!\!\!\!\!\!\!\!\!\!\!\!\!\!\!\!\!\!\!\!\!\!\!\!-\;\mathcal{C}_{k}\;\cheta{\chi}(\kk;b) \sum_{i=1}^{N} \sum_{\alpha}^{x,y,z} 
(U_{i,\alpha}+ V_{i,\alpha})\,\eps_{\alpha}^*\, e^{-i\kk\cdot \RR_i},\label{eq:em32}\\
\hbar\omega\, v_{\kk\epsbs} &=& -\hbar ck\, v_{\kk\epsbs}\notag\\
&&\!\!\!\!\!\!\!\!\!\!\!\!\!\!\!\!\!\!\!\!\!\!\!\!-\;\mathcal{C}_{k}\;\cheta{\chi}(\kk;b) \sum_{i=1}^{N} \sum_{\alpha}^{x,y,z} (U_{i,\alpha}+
 V_{i,\alpha})\,\eps_{\alpha}\, e^{i\kk\cdot \RR_i}\label{eq:em42},
\end{eqnarray}
where 
\begin{multline}
\textrm{d}_B\,{\mathbb E}_{\perp,\alpha}(\RR_i)=\\
-\sum_{\kk\in\mathcal{D}} \sum_{\epsbs\perp\kk} \mathcal{C}_k\left[
\eps_{\alpha}\; u_{\kk\epsbs}\; e^{i\kk\cdot \RR_i} - 
\eps_{\alpha}^*\; v_{\kk\epsbs}\; e^{-i\kk\cdot \RR_i} \right]\cheta{\chi}(\kk;b),
\label{eq:efuv}
\end{multline}

Equations (\ref{eq:em12})-(\ref{eq:em42}) can be rewritten, after simple manipulations, in the useful form
\begin{eqnarray}
U_{i,\alpha} + V_{i,\alpha} &=& \frac{2\omega_B}{\hbar(\omega_B^2-\omega^2)}\;\textrm{d}_B\; {\mathbb E}_{\perp,\alpha}(\RR_i), \label{eq:em13}\\ 
U_{i,\alpha} - V_{i,\alpha} &=& \frac{\omega}{\omega_B}\; (U_{i,\alpha} + V_{i,\alpha}), \label{eq:em23}
\end{eqnarray}
\begin{eqnarray}
u_{\kk\epsbs} &\!\!=\!\!& \frac{\mathcal{C}_k\;\cheta{\chi}(\kk;b)}{\hbar(ck-\omega)} \sum_{i=1}^{N} \sum_{\alpha}^{x,y,z} 
(U_{i,\alpha}+ V_{i,\alpha})\,\eps_{\alpha}^*\, e^{-i\kk\cdot \RR_i},\label{eq:em33}\\
v_{\kk\epsbs} &\!\!=\!\!& - \frac{\mathcal{C}_k\;\cheta{\chi}(\kk;b)}{\hbar(ck+\omega)}\sum_{i=1}^{N} \sum_{\alpha}^{x,y,z} 
(U_{i,\alpha}+ V_{i,\alpha})\,\eps_{\alpha}\, e^{i\kk\cdot \RR_i}.\label{eq:em43}
\end{eqnarray}
Eq.(\ref{eq:em13}) is valid for $\omega_B^2-\omega^2\neq 0$,
and  Eq.(\ref{eq:em33}) is valid for $ck-\omega\neq 0$.
We note that a solution with $\omega=ck$ is possible if one finds a free field
solution (that is a solution of Maxwell equations of frequency
$\omega$ in the absence of charge and current)
such that the electric field ${\mathbb E}_{\perp}$ vanishes
in all atomic locations $\RR_i$, so that all $U_{i,\alpha}+V_{i,\alpha}$ are zero.
This type of free field solutions, already considered in \cite{Knoester06},
are thus left out by Eq.(\ref{eq:em33}) and have to be investigated
by a specific calculation.

\subsection{Periodic system: Bravais lattice}
We now consider $N$ atoms in a generic Bravais lattice, within the quantization volume 
$\textrm{V}=\{\rr\;|\;\rr=\sum_\gamma x_\gamma \mathbf{e}_\gamma,
 0\leq x_\gamma < M_\gamma\}$, $\gamma\in\{1,2,3\}$, $M_\gamma\in\mathbb{N}^*$,
compatible with the lattice geometry.
Thanks to the Bloch theorem one has
 \begin{eqnarray}
U_{i,\alpha} &=& U_{0,\alpha}\;\;e^{i\qq\cdot\RR_i},\label{eq:b1}\\
V_{i,\alpha} &=& V_{0,\alpha}\;\;e^{i\qq\cdot\RR_i},\label{eq:b2}
\end{eqnarray}
where $\qq\in \mathcal{D}$ is the Bloch wavevector in the first Brillouin zone in the reciprocal lattice ($\textrm{RL}$)
and atom $0$ is placed at the origin of coordinates. 
By substituting (\ref{eq:b1}) and (\ref{eq:b2}) in the expressions for $v_{\kk\epsbs}$ and $u_{\kk\epsbs}$ 
of Eqs.(\ref{eq:em33}) and (\ref{eq:em43}), and using the fact 
that the sum $\sum_{i=1}^{N}\;e^{\pm i(\kk\pm\qq)\cdot\RR_i}=0$ for all values of $\kk\pm\qq$ 
that are not points of the reciprocal lattice, 
while $\sum_{i=1}^{N}\;e^{\pm i(\kk\pm\qq)\cdot\RR_i}=\textrm{V}/\mathcal{V}_\textrm{L}$ 
for $\kk\pm\qq=\KK\in\textrm{RL}$, where $\textrm{V}$ is the quantization volume occupied by the $N$ atoms 
and $\mathcal{V}_\textrm{L}$ is the volume of the primitive unit cell in the direct lattice $\textrm{L}$, 
one has that $u_{\kk\epsbs}=0$ and $v_{\kk\epsbs}=0$ except if $\kk=\qq+\KK$, $\KK\in\textrm{RL}$, in which case
\be
u_{\kk\epsbs} = \frac{\textrm{V}}{\mathcal{V}_\textrm{L}}\;\frac{\mathcal{C}_k\;\cheta{\chi}(\kk;b)}{\hbar(ck-\omega)}\; \sum_{\alpha}^{x,y,z} 
(U_{0,\alpha}+ V_{0,\alpha})\,\eps_{\alpha}^*,\label{eq:em34}
\ee
or if $\kk=-\qq+\KK$, $\KK\in\textrm{RL}$, in which case
\be
v_{\kk\epsbs} = - \frac{\textrm{V}}{\mathcal{V}_\textrm{L}}\;\frac{\mathcal{C}_k\;\cheta{\chi}(\kk;b)}{\hbar(ck+\omega)}\; \sum_{\alpha}^{x,y,z} 
(U_{0,\alpha}+ V_{0,\alpha})\,\eps_{\alpha}.\label{eq:em44}
\ee
Inserting (\ref{eq:em34}) and (\ref{eq:em44}) in the expression of ${\mathbb E}_{\perp,\alpha}(\RR_i)$ of Eq.(\ref{eq:efuv}), 
using the relation $\sum_{\epsbs\perp\kk} \eps_{\alpha}\eps_{\beta}^*=\delta_{\alpha\beta}-k_\alpha k_\beta / k^2$, 
and Eq.(\ref{eq:em13}), and using the fact that $\tilde\chi(\kk;b)$
is an even function of $\kk$, we obtain that the eigenfrequency $\omega$ 
is given by the solution of the equation $\det M(b) = 0$
where the matrix elements $M_{\alpha\beta}(b)$ are given in Eq.(\ref{eq:Mfin}).

\subsection{Periodic system: non-Bravais lattice}

Finally, we briefly consider the case of a generic non-Bravais lattice of
section \ref{sec:nonBravais}: in the elementary unit cell, there are
$P$ atoms of positions $\rr_\nu$, $\nu\in \{1,\ldots, P\}$, and this base
is repeated periodically according to an underlying Bravais lattice.
In the equations of subsection \ref{subsec:eqom}
one then has to replace the Bravais lattice positions $\RR_i$ by the atomic positions
$\rr_i$ in the crystal:
\be
\rr_i  = \rr_{\nu_i} + \RR_i,
\ee
where $\RR_i$ belongs to the underlying Bravais lattice and $\rr_{\nu_i}$ is
a position within the elementary unit cell.
Then Bloch theorem gives
\bea
U_{i,\alpha} &=& U_{0,\alpha}^{(\nu_i)} \, e^{i\qq\cdot\RR_i} \\
V_{i,\alpha} &=& V_{0,\alpha}^{(\nu_i)} \, e^{i\qq\cdot\RR_i} ,
\eea
where the Bloch vector $\qq\in \mathcal{D}$ may be chosen in the first Brillouin
zone of the reciprocal lattice RL.
For $\kk=\qq+\KK$, where $\KK \in\mathrm{RL}$,
\be
u_{\kk\epsbs} = \frac{\textrm{V}}{\mathcal{V}_\textrm{L}}\;\frac{\mathcal{C}_k\;\cheta{\chi}(\kk;b)}{\hbar(ck-\omega)}\; 
\sum_{\alpha}^{x,y,z} \sum_{\nu=1}^{P}
(U_{0,\alpha}^{(\nu)}+ V_{0,\alpha}^{(\nu)})\,e^{-i\kk\cdot \rr_\nu} \eps_{\alpha}^*,
\ee
otherwise $u_{\kk\epsbs}$ is equal to zero.
Similarly, for $\kk=-\qq+\KK$, where $\KK \in\mathrm{RL}$,
\be
v_{\kk\epsbs} = -\frac{\textrm{V}}{\mathcal{V}_\textrm{L}}\;\frac{\mathcal{C}_k\;\cheta{\chi}(\kk;b)}{\hbar(ck+\omega)}\; 
\sum_{\alpha}^{x,y,z} \sum_{\nu=1}^{P}
(U_{0,\alpha}^{(\nu)}+ V_{0,\alpha}^{(\nu)})\,e^{i\kk\cdot \rr_\nu} \eps_{\alpha},
\ee
otherwise $v_{\kk\epsbs}$ is equal to zero.
Then proceeding as for the Bravais case one gets (\ref{eq:Mfinveramultieq}).


\begin{thebibliography}{99}

\bibitem{Hopfield58} 
J.J. Hopfield, Phys. Rev. {\bf 112}, 1555 (1958).	

\bibitem{Agranovich60}
V. Agranovich, Sov. Phys. JETP {\bf 37}, 307 (1960).

\bibitem{Fano56} 
U. Fano, Phys. Rev. {\bf 103}, 1202 (1956).

\bibitem{LagendijkRMP}
P. de Vries, D.V. van Coevorden, A. Lagendijk, Rev. Mod. Phys. {\bf 70}, 447 (1998).

\bibitem{Juzeliunas07} 
J. K\"astel, M. Fleischhauer, and G. Juzeli\={u}nas, Phys. Rev. A {\bf 76}, 062509 (2007),
and references therein.

\bibitem{Carusotto08}
I. Carusotto,
M. Antezza, F. Bariani, S. De Liberato, and C. Ciuti
Phys. Rev. A {\bf 77}, 063621 (2008).

\bibitem{Lagendijk96} 
D.V. van Coevorden, R. Sprik, A. Tip, and A. Lagendijk, Phys. Rev. Lett.  {\bf 77}, 2412 
(1996).

\bibitem{Knoester06} 
J.A. Klugkist, M. Mostovoy, and J. Knoester, Phys. Rev. Lett. {\bf 96}, 163903 (2006).

\bibitem{Bloch02} 
M. Greiner, O. Mandel, T. Esslinger, T.W. H\"{a}nsch, and I. Bloch, 
Nature {\bf 415}, 39 (2002).

\bibitem{Grynberg94}
K. I. Petsas, A. B. Coates, and G. Grynberg, Phys. Rev. A
{\bf 50}, 5173 (1994).

\bibitem{Soukoulis90} 
K.M. Ho, C.T. Chan, and C.M. Soukoulis, Phys. Rev. Lett. {\bf 65}, 3152 (1990).

\bibitem{FCBook}  
J.D. Joannopoulos, S.G. Johnson, J.N. Winn, R.D. Meade, 
\emph{Photonic Crystals: Molding the Flow of Light}, 2nd ed. (Princeton Univ. Press, 2008).

\bibitem{Yablo91}
E. Yablonovitch, T.J. Gmitter, K.M. Leung, Phys. Rev. Lett.
{\bf 67}, 2295 (1991).

\bibitem{Castin09} M. Antezza and Y. Castin, arXiv:0903.0765v1 (2009).

\bibitem{John04}  %Photonic Band Gap Architectures for Holographic Lithography
O. Toader, T.Y. Chan, and S. John, Phys. Rev. Lett. {\bf 92}, 043905 (2004).

\bibitem{dupdc}
The use of a Gaussian cut-off ensures a rapid convergence 
of the sums to appear over the reciprocal lattice.
Combined with the relations (\ref{eq:d0}) and (\ref{eq:domulti}), it allows to avoid
Ewald summation tricks \cite{Mahan65,Philpott73}, with which it shares however
some mathematical features.

\bibitem{Mahan65} 
G.D. Mahan, J. Chem. Phys. {\bf 43}, 1569 (1965).

\bibitem{Philpott73}
M.R. Philpott, J.W. Lee, J. Chem. Phys. {\bf 58}, 595 (1973).

\bibitem{box}
Not all cut-off functions lead to an expression for (\ref{eq:Mfin}) 
which is possible to renormalize when $b\to 0$.
Surprisingly, even by considering the simple case of a simple cubic lattice
of lattice constant $a$,
if one chooses for $\tilde{\chi}(\kk;b)$ the function equal to one
for $\kk$ inside the cubic box 
$[-\frac{\pi}{b},\frac{\pi}{b}]^3$ and 
zero elsewhere, $a/b$ being integer,
one obtains that,  when $b\to 0$, the sum (\ref{eq:Mfin}) diverges, 
and, most important, the divergent terms depend on the Bloch wavevector $\qq$, 
hence providing a non-renormalizable model. 
To show this,  let us consider the sum  (\ref{eq:Mfin}) for $\alpha=x$ and $\beta=y$. 
By using symmetry properties of the cubic geometry the sum has, for $b\to 0$,
a leading term
\[
q_x q_y \;\times
\sum_{\KK\in\textrm{RL}\cap[-\frac{\pi}{b},\frac{\pi}{b}]^3} 
\;\left(\frac{8K_x^2K_y^2}{K^6}-\frac{1}{3K^2}\right)\;
\label{eq:div}
\]
that diverges in a way proportional to $q_x q_y$ for $b\to 0$.

\bibitem{Ryder}
L.H. Ryder, in {\sl Quantum Field Theory} (Cambridge University Press, 1996).

\bibitem{conditions}
For $\omega_0^2$ to be positive, $b$ should be large enough.
For an hydrogen-like atom we may use the estimates $\hbar\omega_B \approx q^2/(4\pi\varepsilon_0 a_0)$
and $\textrm{d}_B\approx |q| a_0$, where $q$ is the electron charge
and $a_0$ the atomic radius,
so that $\textrm{d}_B^2/(\hbar \varepsilon_0 \omega_B)\approx 4\pi a_0^3$.
The first factor in (\ref{eq:ren1}) is positive if  $a_0 \lesssim b$,
which is natural in a dipolar coupling model.
Then the second factor is close to unity if
$(a_0/b)\times (\omega_B a_0/c)^2 \lesssim 1$.
This condition is satisfied, 
since $\omega_B a_0/c \approx \alpha$, where $\alpha\simeq 1/137\ll 1$ 
is the fine structure constant.

\bibitem{demons}
The electric field of a free-field solution is a sum of plane waves
which interfere destructively in all atomic positions.
This sum thus includes at least two plane waves. 
Let us select arbitrarily two of these plane waves; if one has
a wavevector $\kk$, the other one has a wavevector necessarily
of the form $\kk+\KK$, where $\KK \in \textrm{RL}^*$, because of the Bloch theorem.
These two waves have the same frequency $\omega_{\rm free}$ and obey the free space
dispersion relation, so that $\omega_{\rm free}/c= k = |\kk+\KK|$.
Squaring this last equality gives $2\kk\cdot\KK + K^2=0$, so that
$\kk=\kk_{\perp} - \KK/2$ where $\kk_{\perp}$ is orthogonal to $\KK$.
Then $k\geq K/2$, which leads to (\ref{eq:borne}).

\bibitem{derivation}
One introduces $S_{\alpha\beta}(\kk;b)= \sum_{\RR\in L^*} \exp(i\kk\cdot\RR)
\langle u_{\alpha\beta}(\rr+\RR)\rangle_{\rr}$, with $u_{\alpha\beta}(\rr)=
(\delta_{\alpha\beta}-3r_\alpha r_\beta/r^2)/(4\pi r^3)$ and $\langle \ldots \rangle_{\rr}$
denotes the average over $\rr$ with the Gaussian probability distribution
$\propto \exp(-r^2/4 b^2)$. Using Poisson's formula and restricting to 
$k\ll\inf_{\mathbf{K}\in RL^*} K$,
one obtains $\lim_{b\to 0} S_{\alpha\beta}(\kk;b)\simeq \left[\frac{k_\alpha k_\beta}{k^2}
-\delta_{\alpha\beta} + \textrm{J}_{\alpha\beta}\right]/\mathcal{V}_L$.
The Fourier transform of $u_{\alpha\beta}$ is indeed
$\tilde{u}_{\alpha\beta}=\frac{k_\alpha k_\beta}{k^2}-\frac{1}{3} \delta_{\alpha\beta}$.
We define the right-hand side of (\ref{eq:electro}) 
as $\lim_{b\to 0} \langle S_{\alpha\beta}(\kk;b)\rangle_{\kk}$, where $\langle \ldots \rangle_{\kk}$
stands for the uniform average over the direction of $\kk$. We then get (\ref{eq:electro}).

\bibitem{question}
A mathematical question is to know what is the exact minimal free-field frequency 
in the diamond atomic lattice. Within the two-mode ansatz of \cite{Knoester06} 
one finds a minimal $\omega_{\rm free}/c=2\pi\sqrt{2}/a$.
The general three-mode ansatz can give lower frequency solutions that however
form a discrete set. The corresponding minimal value is
$\omega_{\rm free}/c=3\pi/(a\sqrt{2})$;  it is obtained by superimposing
with the right amplitudes the three plane waves of wavevectors
$\kk=(-3\pi/2a,3\pi/2a,0)$, $\kk-\tilde{\mathbf{e}}_1$ and $\kk+\tilde{\mathbf{e}}_2$,
with a common polarization $\epsb=(1,1,0)$.
%Of course $3/\sqrt{2} > \sqrt{3}$.
We have not explored the ansatz with four waves or more.

\bibitem{David02}
Y. Bidel, B. Klappauf, J.-C. Bernard, D. Delande, G. Labeyrie, 
C. Miniatura, D. Wilkowski, R. Kaiser,
Phys. Rev. Lett. {\bf 88}, 203902 (2002).

\bibitem{Yabu03}
Y. Takasu, K. Maki, K. Komori, T. Takano, K. Honda,
M. Kumakura, T. Yabuzaki, and Y. Takahashi, Phys.
Rev. Lett. {\bf 91}, 040404 (2003).

\bibitem{Takahashi09}  
T. Fukuhara, S. Sugawa, M. Sugimoto, S. Taie, Y. Takahashi, 
Phys. Rev. A {\bf 79}, 041604 (2009).

\bibitem{Phillips07}
M. Anderlini, P.J. Lee, B.L. Brown, J. Sebby-Strabley,
W.D. Phillips, J.V. Porto, Nature {\bf 448}, 452 (2007).

\bibitem{Courtois99}
A. Chelnokov, S. Rowson, J.-M. Lourtioz, V. Berger,
J.-Y. Courtois, J. Opt. A: Pure Appl. Opt. {\bf 1}, L3 (1999).

\bibitem{maxima}
The locations of the intensity maxima are $\pm (a/8,a/8,a/8)$
modulo any vector of the fcc lattice, so that they also form
a diamond lattice.

\end{thebibliography}
\end{document}